\documentclass[preprint]{aastex}
\usepackage{bm}

\usepackage{color}
\usepackage{graphicx}
\newcommand{\be}{\begin{equation}}
\newcommand{\ee}{\end{equation}}
\newcommand{\cov}{\mathrm{Cov}}

\newcommand{\ba}{\mbox{\boldmath{$\alpha$}}}
\newcommand{\bt}{\mbox{\boldmath{$\theta$}}}

\slugcomment{ANL-HEP-PR-12-3}
\begin{document}

\title{ Mass Reconstruction using Particle Based Lensing II: Quantifying substructure with Strong+Weak lensing and X-rays.}
\author{Sanghamitra Deb} 
\affil{Argonne National Laboratory, 9700 South Cass Avenue, Lemont, IL 60439, USA ,  Lawrence Berkeley National Laboratory,  Drexel University. } 
\email{sdeb@hep.anl.gov}
\author{Andrea Morandi}
\affil{School of Physics and Astronomy,
Tel Aviv University,
Tel Aviv, 69978 }
\author{Kristian Pedersen}
\affil{Dark Cosmology Centre, Niels Bohr Institute and Space Science Center, University of Copenhagen.} 
 \author{Signe Riemer-Sorensen}
\affil{Dark Cosmology Centre, Niels Bohr Institute,  School of Mathematics and Physics, University of
Queensland, QLD 4072, Australia.} 
 \author{David M. Goldberg}
 \affil{Department of Physics, Drexel University, Philadelphia PA-19104}
\author{H{\aa}kon Dahle}
\affil{Institute of Theoretical Astrophysics, University of Oslo. P.O. Box 1029, Blindern, N-0315 Oslo, Norway}

\begin{abstract}
We report a mass reconstruction of A1689 using Particle Based Lensing (PBL), a new technique for Strong+Weak lensing that allows a variable resolution depending on the data density and the signal-to-noise. Using PBL we also calculate the covariance matrix for the resulting mass map.
  The reconstruction of A1689 shows a secondary mass peak in the north-east direction confirming previous optical observations. This indicates that the central region of the cluster is still undergoing a weak merger. We have used this mass map to measure power ratios of the dark matter distribution and compared it to the X-ray distribution. We find that the power in the X-ray distribution is lower suggesting a smoother and rounder gas distribution compared to the dark matter distribution. We fitted an NFW profile to the profile derived from the mass map and we find that the lensing mass within 1 Mpc is $1.5\pm0.33\times 10^{15}M_\odot$. This is higher than the X-ray mass.
\end{abstract}
  

\maketitle

\section{Introduction}
Galaxy clusters are the largest gravitationally bound objects in the universe. They produce distinct observational signatures (strong emission in the x-ray and sub-millimeter wavelengths, distorted background galaxies) providing information that constrain cosmological structure formation, dark matter dynamics, gas-physics  and galaxy formation.  The number of galaxy clusters as a function of mass and redshift, i.e the mass function of clusters will provide strong constraints on stucture formation models, the normalization of the matter power spectrum, cosmological matter density $\Omega_m$, and evolution of dark energy $w(z)$ \citep[eg.][]{2009MNRAS.393L..31F,2006A&A...454...27B,1996MNRAS.282..263E,1999MNRAS.308..119S,1990ApJ...351...10F,1974ApJ...187..425P}. Cluster mass measurements are usually plagued by assumptions related to hydrostatic equilibrium and dynamics of the galaxies. Gravitational lensing measures the mass of the clusters without making any assumptions about the dynamical state of the cluster. The primary bias in this case is from line-of-sight projection effects \citep{2011arXiv1111.0989S, 2011MNRAS.416.1392G} which is small for very high mass clusters \citep{2011arXiv1106.2046B}.


Galaxy clusters are still in the process of formation and consequently a large fraction of them have secondary structure. The presence of substructure reduces the accuracy of techniques that assume the entire mass to be associated with a single peak. Hence it is important to quantify substructure from lensing and X-ray measurements. 
 \cite{2008ApJ...685..118V} have fit an empirical dependence of mass on substructure parameters and \cite{2008ApJ...681..167J} have shown a strong correlation between inaccurate hydrostatic masses and presence of substructure from simulations.  

Gravitational lensing not only constrains the projected mass of the cluster, it gives us information about the distribution of dark matter. Weak lensing analysis has produced interesting results like the cosmic train wreck A520 \citep{2007ApJ...668..806M}, the supercluster A901/902 \citep{2008MNRAS.385.1431H,2010ApJ...721..124D} and many others \citep{2009arXiv0903.1103O, 2007MNRAS.374L..37K,2007ApJ...667...26P,2001AJ....121...10C,1999ApJ...527..535W,1998MNRAS.296..392A}. 
Strong lensing \citep{2010MNRAS.408.1916Z, 2010arXiv1009.3936Z} lets us resolve the core of the cluster in finer detail providing additional information on substructure properties at the core. The combination of weak and strong lensing lets us probe the mass distribution both at the core and the periphery of the cluster. In this paper we will reconstruct the mass map of Abell 1689 using both weak and strong lensing.



Abell 1689 is an X-ray luminous cluster with a velocity dispersion as high as $\sim|1600|$ km/sec~\citep{2009ApJ...701.1336L} at a redshift of 0.18 and it is one of the richest clusters observed to date. It is well known for its spectacular arcs and many multiple images. 
There have been several lensing studies of A1689 from various facilities ranging from SUBARU, CFHT, HST and many others \citep{2005ApJ...619L.143B} and different groups  have found a wide range of values for the measured concentration of the cluster \citep[Table 4,][hereafter L07]{2007ApJ...668..643L}. Strong lensing reconstruction using both parametric 
(L07) and non-parametric methods \citep[][hereafter C10]{2010arXiv1005.0398C}  have also estimated a higher concentration (compared to expectation from $\Lambda$CDM simulations) for an NFW profile fitting. \cite{2005MNRAS.362.1247D} have done a multi-resolution non-parametric mass reconstruction for A1689 using strong lensing only for the core of the cluster.

In this paper we will develop a Strong+Weak (S+W) lensing analysis using Particle Based Lensing \citep[PBL][] {2008ApJ...687....1D, 2010ApJ...721..124D}. The advantage of using PBL is the variable resolution that can be obtained in the strong (high resolution) and weak(low resolution) lensing regions. Additionally, the errors in this technique are well understood. This makes calculation of moments from the mass map possible. We apply the S+W lensing with PBL to A1689 and compare the mass profile and the substructure parameters from lensing and X-ray measurements.

\section{Lensing Basics}
The lensing notation followed in this paper will be similar to  \cite{2008ApJ...687....1D, 2010ApJ...721..124D}. The ultimate goal is to map the two-dimensional dimensionless surface mass density $\kappa$ 
The lensing fields, namely dimensionless surface mass density $\kappa$, the two components of shear $\gamma^{1,2}$. The components of the deflection field $\alpha^{1,2}$  is related to the potential $\psi$ by,

\begin{eqnarray}
\kappa=\frac{\psi_{11}+\psi_{22}}{2}\\
\gamma^1=\frac{\psi_{11}-\psi_{22}}{2}\\
\gamma^2=\psi_{12}\\
\alpha^{1,2}=\psi_{1,2}
\end{eqnarray}

where $\psi_{ij}$ represents $\frac{\partial^2\psi}{\partial x_i \partial x_j}$ and $\psi_i$ refers to $\frac{\partial \psi}{\partial x_i}$.

Using PBL the $\kappa$ is related to the potential via a linear matrix operation $\hat\kappa$ given by,
\begin{eqnarray}
\kappa=\hat{\mathrm{K}} \psi
\label{eqn:pbl}
\end{eqnarray}

Similar relations exist for the shear and the deflection field.
The redshift dependence of the lensing fields : $\kappa,\gamma, \alpha$ is incorporated by multiplying them with the redshift weight factor given by,

\begin{equation} 
Z(z)=\frac{(D_{ls}/D_s)(z)}{(D_{ls}/D_{s})(z\to\infty)}
\end{equation}
The redshift weight factor plays an important role for breaking the mass sheet degeneracy with strong lensing data.
The weak lensing observables are ellipticities of the background galaxies. In the weak lensing regime ($\kappa<1$) the ellipticity is given by,
\begin{equation}
\langle\varepsilon^{i}\rangle=g^i=\frac{\gamma^{i}Z(z)}{(1-\kappa Z(z))}
\label{eqn:ellipticity1}
\end{equation}

In the strong lensing regime defined by $((1-\kappa)^2-|\gamma|^2) < 0$, the ellipticity $\langle\varepsilon^{i}\rangle$ is given by,

\begin{equation}
\langle\varepsilon^{i}\rangle=g^i=\frac{(1-\kappa Z(z))\gamma^{i}Z(z)}{|\gamma Z(z)|^2}
\label{eqn:ellipticity2}
\end{equation}

The strong lensing observables are the angular positions of the multiply imaged sources. The difference of the angular distance between a pair of images is directly related to the deflection angle via,

\begin{equation}
\theta^A-\theta^B=\alpha^A Z(z)-\alpha^B Z(z)
\label{eqn:strong}
\end{equation}

where $A$ and $B$ represent the two lensed images from the same source. While combining the weak and the strong lensing information we use Equations ~\ref{eqn:ellipticity1}, \ref{eqn:ellipticity2} and~\ref{eqn:strong} simultaneously with ~\ref{eqn:pbl} to map the dimensionless surface mass density $\kappa(\theta,z)$.

\section{Strong and Weak lensing by Galaxy clusters}
\label{sec:ws}
In regions with $\kappa>1$ rays of light from distant sources take multiple paths from the source to reach the observer producing multiple images. There are certain configurations of lens and source galaxies that give rise to giant arcs and multiple images making them very rare. The accuracy of measuring strong lensing constraints is very high, however since they are rare and occur only in the core of the cluster we need weak lensing to constrain the mass distribution in the outer parts of the cluster. Weak lensing causes subtle distortion of background galaxies in regions where $\kappa<1$.
 
The primary challenge in combining strong and weak lensing data is the difference in scales at which the various signals dominate. Since galaxies are intrinsically elliptical and the distortion due to lensing is  at the level of a few percent , several tens of galaxies have to be averaged to extract a weak lensing signal. This limits the resolution of  weak lensing mass reconstructions to vary from $1^\prime$ for ground based data to $0.5^\prime$ for space based data. On the other hand strong lensing occurs within the central one arcminute region of the cluster. The detailed information about the structure of the cluster core will be smoothed out if the combined mass reconstruction is done with the weak lensing resolution. This situation requires a mass reconstruction technique with variable resolution. PBL is designed for this purpose, the resolution of the reconstructed mass is a function of the local number density of data and the signal-to-noise of the observables. Thus using this technique we get a higher resolution at the core and a lower resolution at the periphery of the cluster.

 A combined likelihood function is given by,
\begin{eqnarray}
\label{eqn:chi0_0}
{\cal L}&\propto&\exp\left[-\frac{\sum_{mn} \left({\varepsilon}^{i}_m-\frac{\gamma^i_m\{\psi_m\}}{1-\kappa_m\{\psi_m\}}\right)C^{-1}_{mn} \left({\varepsilon}^{i}_n-\frac{\gamma^i_n\{\psi_n\}}{1-\kappa_n\{\psi_n\}}\right)}{2}\right.\\ \nonumber
&&\left.-\sum_{i,pairs}\frac{\left((\ba^A(\{\psi\})-\ba^B(\{\psi\}))Z(z_i)-(\bt^A-\bt^B)\right)^2}{\sigma_i^2}\right].
\end{eqnarray}
Here $n,m$ refer to the particle index.
The first term of the equation is due to weak lensing only. The covariance in the data arises because of the smoothing procedure described in \cite{2010ApJ...721..124D}. The second term represents the fit between the multiple images and  the deflection field at the location of the multiple images scaled by the magnification at the image plane.  Maximizing the liklihood for anyone of these terms is simple. The weak lensing mass map has a resolution of $0.5^\prime$, and the strong lensing mass map can have a resolution as low as $10^{\prime\prime}$~\citep{2010arXiv1005.0398C} and the positions of the multiple images are fit exactly. The Strong+Weak (S+W) reconstruction has a variable resolution, the outskirts have a resolution determined by the smoothing scale of the weak lensing data and the core has a higher resolution since we fit to the high signal-to-noise strong lensing data. The simultaneous fitting of the strong and weak lensing data is complicated since there is some ambiguity in the choice of relative weighting between the strong and the weak lensing measurements. We weight the strong lensing constraints with the inverse of the effective error in Equation~\ref{sig_sl}. The advantage of an S+W reconstruction is that it ensures the mass map at the core of the cluster is consistent with the outskirts.

The contribution to the error of the $i^{th}$ pair is given by a combination of the error in redshift and the astrometric error in measuring the positions of the multiple images. The error in measuring the positions of the images is $\sigma_\theta=0.2^{\prime\prime}$ (L07). The error $\sigma$ is given by,
\be
\sigma^2=\left(\frac{\Delta \theta}{Z(z)}\right)^2\left(\frac{\sigma_\theta}{\Delta \theta}\right)^2+\left(\frac{\Delta \theta }{Z(z)}\right)^2\left(\frac{\sigma_{Z(z)}}{Z(z)}\right)^2.
\label{sig_sl}
\ee
where $\Delta\theta$ is the difference in positions for multiple images, $\sigma_\theta$ are the astrometric errors and $\sigma_z$ are the errors in redshift.

The astrometric errors associated with the positions of the multiple images is low. However the error in the strong lensing mass reconstruction is dominated by Poisson error. Multiple images sample the cluster potential at discrete locations at finite number of points. Thus, even if we fit the mass at those finite locations very accurately the overall mass distribution will have higher poisson error. The PBL reconstruction also has higher residuals (Figure~\ref{fig:resi}) for the strong lensing images compared to C10 in order to incorporate the weak lensing constraints which have higher errors due to intrinsic ellipticity of source galaxies. 

\subsection{Covariance of S+W map}
The minimum of the $\chi^2=-2 \mathrm{log}({\cal L})$ (defined in Equation~\ref{eqn:chi0_0}) gives the best solution for the potential $\psi$ and hence the mass. In order to calculate the error in the mass distribution corresponding to an error in the observables we differentaite the $\chi^2$ at its minimum and get,

\begin{equation}
\sum_{i}(G^{i})^{T}C^{-1} G^{i}\psi+w_s\sum_{i}A^{i,T}A^{i}\psi=\sum_i(G^{i})^{T}C^{-1} \hat{ \varepsilon}^{i}+w_s \sum_i A^{i,T} \delta_x^{i}.
\label{eqn:psi}
\end{equation}
Here $\hat\varepsilon$ is the smoothed ellipticity defined in \cite{ 2010ApJ...721..124D}, and $\delta_x$ are the residuals to the fit of the strong lensing observables.

The solution for the potential is obtained by inverting the above equation,
\be
\psi=V_w \hat\varepsilon +V_s \delta_x,
\ee
where 
\be
V_w=\left(\sum_{i}(G^{i})^{T}C^{-1} G^{i}+w_s\sum_{i}A^{i,T}A^{i}\right)\sum_i(G^{i})^{T}C^{-1},
\ee
and 
\be
V_s=\left(\sum_{i}(G^{i})^{T}C^{-1} G^{i}+w_s\sum_{i}A^{i,T}A^{i}\right)w_s \sum_i A^{i,T}.
\ee

The covariance in the potential is given by,

\be
\cov^\psi =\psi \psi^{T}=V_w \cov^{\varepsilon} V_w^{T}+V_s \cov^{\delta_x} V_s^{T}.
\label{eqn:covcomb}
\ee

Here we have assumed that the terms $V_w\hat\varepsilon\delta_x^T V_x^T$ and $V_s\delta_x\hat\varepsilon^{T}V_w^{T}$ go to zero since 
\be
\langle\gamma\alpha\rangle=0
\ee
The above relationship holds since the shear has $m=2$ symmetry and the deflection field has an $m=1$ symmetry. 
Additionally there is very little overlap between weak and strong lensing data since the central region of clusters is masked by the presence of bright cluster member galaxies. This ensures that there is no correlation between the weak and the strong lensing data.
The covariance in Equation~(\ref{eqn:covcomb}) has been calculated in~\cite{2010ApJ...721..124D}.

The noise covariance matrix for the reconstructed $\kappa$ map is given by,
\be
\cov^\kappa(\zeta)=K \cov^{\psi} K^{T} ,
\ee
 Here the reconstructed mass is dependent on the scale $\zeta$ used to smooth the ellipticities. It is important to note that the resolution scale in the strong lensing regime is dictated by the separation between the multiple images.

\section{Data}
\subsection{Weak Lensing}
The weak gravitational lensing measurements used in this study were based on deep archival imaging data of A1689 from the Suprime-cam wide-field 
imager \citep{2002PASJ...54..833M} at  Subaru telescope. Data representing 
a total exposure time of 1920s in the $V-$band and 2640s in the SDSS $i^\prime-$band were retrieved from the SMOKA archive system and reduced using the standard SDFRED reduction pipeline for Suprime-cam  data \citep{2002AJ....123...66Y}. The limiting magnitude of the Subaru data are AB magnitude V=26.5 and $i^\prime=25.9$ for a $3\sigma$ detection within a $2^{\prime\prime}$ aperture. 

Following \cite{2005ApJ...619L.143B}, the $V-i^\prime$  color information from the Suprime-cam photometry was used to separate gravitationally 
lensed background galaxies from foreground and cluster galaxies by selecting galaxies that have $V-i^\prime$ color $> 0.22$ magnitudes 
redder than the red sequence formed by early-type cluster galaxies in a color-magnitude diagram. This catalogue was augmented by a partially 
overlapping set of background galaxies, selected to have an estimated photometric redshift larger than $z=0.4$, 
calculated from the CFH12k $BRI$ imaging data of  \cite{2005A&A...434..433B}. The union of these two background galaxy samples,
with an $i^\prime-$band  $S/N > 7$, consists of 9943 objects within a sky area of 33'$\times$23'. 
Our lensing distortion measurements, based on this galaxy set, made use of the IMCAT software package \citep{1995ApJ...449..460K, 1997ApJ...475...20L, 1998ApJ...504..636H} with an implementation tested against the STEP simulations of weak lensing data \citep{2006MNRAS.368.1323H}.

\subsection{Strong Lensing}
The strong lensing data used in this analysis is tabulated in L07. 
In order to identify all the strong lensing images a composite image of HST/ACS data in the four bands F475W, F625W and F775W and F850LP are necessary. The accuracy of the modeling of the strong lensing constraints depends on the knowledge of the redshifts of the background sources. The first strong lensing analysis for this cluster was done by \cite{2005ApJ...621...53B}(hereafter B05). In B05 about half of the multiply imaged galaxies were identified by eye to create a mass model which was then used to detect more multiple images. There was misidentification of a few of the multiple images which were corrected in L07 and C10 and additional multiple image systems have been identified. Typically one (or two in some cases) of the multiple images have spectroscopically confirmed redshifts, the other members of that particular system are obtained by fitting a model and ensuring that they have the same color.

\subsection{X-ray Data}
Clusters of galaxies show very strong X-ray emission. Due to the low efficiency of galaxy formation, $90\%$ of the baryons are in the form of intergalactic gas. The deep potential well of galaxy clusters traps this gas and heats it  to X-ray emitting temperatures. The X-ray temperature serves as a proxy for the depth of the potential well and hence the mass of the cluster.  
The X-ray observable is the X-ray surface brightness distribution due to free-free emission. It is proportional to the integral of $n_e^2$ where $n_e$ is the electron density. In presence of substructure clumps can be visible at varying energies and radii. Details on the X-ray can be found in \cite{2011MNRAS.416.2567M,2009ApJ...693.1570R}. 

We use two Chandra X-ray observations (observation ID 6930 and 7289) from the NASA HEASARC archive with a total exposure time of approx. 150 ks. The observations have been carried out using ACIS--I CCD imaging spectrometer. We reduced these observations using the CIAO software (version 4.1.2) distributed by the {\it Chandra} X-ray Observatory Center, by considering the gain file provided within CALDB (version 4.1.3) for the data in VFAINT mode. Brightness images have been extracted from processed event files in the energy range ($0.5-5.0$ keV), corrected by the exposure map to remove the vignetting effects: the two final brightness images has been stacked together, in order to increase the signal-to-noise ratio. All the point sources have been masked out by both visual inspection and the tool {\tt celldetect}, which provide candidate point sources. The background counts have been estimated by selecting regions of the field not contaminated from source emissions. The temperature of the X-ray gas is calculated in radial bins by fitting an absorbed optically-thin plasma emission model
 The deprojected temperature is used to calculate the mass profile using the hydrostatic mass equation \citep{1988xrec.book.....S}.  The resulting mass profile is commonly fitted to the universal NFW profile \citep{1997ApJ...490..493N} to constrain dark matter halo properties. 

\section{Substructure Parameters}

Galaxy clusters that are dynamically young still in the process of formation. Simulations \citep{2008ApJ...672..122E, 2006ApJ...650..128K} indicate that clusters depart from hydrostatic equilibrium at varying degrees suggesting the presence of substructure even in the ``relaxed" systems. The presence of secondary structure leads to inaccuracies in model dependent mass measurements for systems which are very far from equilibrium. Hence it is important to quantify the presence of substructures. In this section we will outline techniques for measuring substructure parameters from observations.  We will calculate two parameters: power ratios and ellipticity of the cluster halo. Simulations have studied the empirical dependence \citep{2008ApJ...685..118V} of mass from X-ray distributions on these parameters.  

\subsection{Power Ratios}
In order to quantify the substructure from non-parametric mass maps we use the simplest form of parameterization. The surface mass density $\Sigma$ is related to the two-dimensional potential:
\be
\nabla^2 \psi=4 \pi \mathrm{G} \Sigma
\ee
 We do a multipole expansion of the potential \citep{1995ApJ...452..522B} given by,

\be
\psi=\psi_0+\psi_1\sum_m\frac{1}{m r^m}\left(a_m \mathrm{cos}(m\phi)+b_m \mathrm{sin} (m\phi)\right)
\ee
where $\psi_0$ and $\psi_1$ are constants and (r,$\phi$) are conventional polar coordinates. Here $a_m$ and $b_m$ are moments of the surface mass density calculated within a circular aperture. The $m^{th}$ moment in the x and the y direction is given by,

\begin{eqnarray}
a_m(r)=\int_{r^{\prime}<r} \Sigma({\vec r}^{\prime})\left(r^{\prime}\right)^m \mathrm{cos}(m\phi^{\prime}) d^2{\vec r^\prime},\\
b_m(r)=\int_{r^{\prime}<r} \Sigma({\vec r}^{\prime})\left(r^{\prime}\right)^m \mathrm{sin}(m\phi^{\prime}) d^2{\vec r^\prime}.
\label{eqn:ambm}
\end{eqnarray}

This technique of multipole expansions is directly related to the dynamical state that results from fluctuations in the cluster potential.
These moments have the following properties. A circularly symmetric mass distribution produces a monopole only term. The dipole term vanishes if the coordinate system is set to be at the center of the mass distribution. An ellipse contributes to even terms only, thus significant contribution to odd terms would indicate presence of substructure. These moments are calculated in a circular aperture. This makes sure that the shape of the aperture does not produce any bias. 

The powers are given by,
\begin{eqnarray}
P_0=[a_0 \ln(R)]^2,\\
P_m=\frac{1}{2 m^2 r^{2m}}\left(a_m^2+b_m^2\right).
\end{eqnarray}
We calculate these powers and calculate their ratio  in the form $P_2/P_0$,$P_3/P_0$ and $P_4/P_0$. These ratios are very sensitive to substructure and describe a wide range of cluster morphologies \citep{2005ApJ...624..606J}. 


\subsection{Ellipticity}
Ellipticity is another measure of secondary structure. Higher ellipticity of halos indicates the presence of sub-peaks in the gas and mass distribution. There is a strong dependence of ellipticity on amplitude of mass fluctuations $\sigma_8$ \citep{2006ApJ...647....8H}. A higher value of $\sigma_8$ indicates that cluster formation has started earlier and hence the measured ellipticity of clusters in the local universe would be lower. 
 Clusters are formed through hierarchical merging of smaller dark matter halos. Thus at their infancy they have more infalling matter and are more elliptical. As they virialise they become more and more spherical. Thus we expect clusters at higher redshift to be more elliptical than low redshift clusters. This has been confirmed by measuring higher order moments of the X-ray gas distribution  \citep{2005ApJ...624..606J,1995ApJ...452..522B}.

The ellipticities are defined as,

\begin{equation}\label{aa3412}
e = \frac{1- \lambda_{-}/\lambda_{+}}{1+\lambda_{-}/\lambda_{+}}
\end{equation}
where $\lambda_{\pm} (\lambda_{+}\ge \lambda_{-})$ are the eigenvalues of the moment of inertia tensor $I$ defined in \cite{2010ApJ...721..124D}.
the position angle of the major axis measured north through east in celestial coordinates can be estimated through the following equation:
\begin{equation}\label{aa3413}
\theta = \arctan{\left({\frac{I_{xx}}{\lambda_{+}^2-I_{yy}}}\right)}+{\frac{\pi}{2}}
\end{equation}

\subsection{Measuring Substructure Parameters}
\noindent{\it Lensing}\\
In \cite{2010ApJ...721..124D} we have discussed calculating moments when the covariance of the mass distribution is known. Here we will give a brief recapitulation, since the $\kappa$ is correlated, we will use singular value decomposition to define a $\kappa^\prime$ that is uncorrelated.

\be
\kappa^\prime=\langle\kappa\rangle+U^T(\kappa-\langle\kappa\rangle)
\ee
where U is the orthogonal matrix from singular value decomposition of the covariance of the $\kappa$ map.
The mean density is given by,
\be
\langle\kappa\rangle=\frac{\sum_{mn} C^{-1}_{mn} \kappa_m}{\sum_{mn}C^{-1}_{mn}}
\ee

 We use this $\kappa^\prime$ as our surface mass density in equations~\ref{eqn:ambm} and inverse weight it by the elements of the diagonal matrix from the singular value decomposition. Thus we make sure that the weighting is done with the uncorrelated $\kappa$.
The the $a_m,b_m$ are weighted by $W=w/s$, where $w$ is gaussian with width given by the radial distance at which the measurements are made. The errors are given by:
\be
\delta_{X}=\sqrt{\frac{\sum(W^2) (X_i-\langle X \rangle)^2}{(\sum W)^2} }
\label{eqn:err}
\ee 
The errors in the power ratio are calculated algebraically from these expressions.
In case of measuring ellipticities  the weight $W=\kappa^\prime w/s$ and the error calculations are similar to Equation~\ref{eqn:err}.

\noindent{\it X-ray}\\
Following \cite{buote1994,2011ApJ...729...37M}, we calculate the eccentricity $e$, centroid and position angle $\theta$ of the X-ray surface brightness distribution. The parameters obtained from this method have been computed within circular regions of increasing radius $R=(200,400,600,800,1000)$ kpc centred on the centroid of the brightness distribution. To determine these parameters from the j$th$ circular aperture of $P_{\rm j}$ pixels having $n_i$ counts in pixel $i$, we calculate the moments:
\begin{equation}\label{aa3411}
\mu_{m,n} = {{\frac{1}{N_j}}  \sum_{i\in \rm{j}th \; \rm{bin}} n_i (x_i-{\stackrel{-}{x}})^m \;(y_i-{\stackrel{-}{y}})^n}\quad  (m,n \le 2), 
\end{equation}
where $N_j=\sum_{i\in  \rm{j}th \; \rm{bin}} n_i$, and $({\stackrel{-}{x}}, {\stackrel{-}{y}})$ is the centroid given by equation $\mu_{1,0}=0$ and $\mu_{0,1}=0$, respectively. We compute $e$ and $\theta$ within circular apertures with increasing radii. Characterization of the errors in this procedure has been performed via Monte-Carlo (MC) randomization of uncertainties due to Poisson statistics of the surface brightness: we performed $10^3$ MC simulations for each of the circular apertures.

For the calculation of the power ratio, we exploit the method as developed by Buote \& Tsai (1995, 1996). The X-ray background has been accounted for by calculating the moments for both the cluster image and the background image (created via poissonian randomization of the background as estimated in regions of the field not contaminated from source emissions) and then subtracting the background moments from the observed moments. 
We compute power ratios within circular apertures with increasing radii $R$. 

\section{Results}
\begin{figure*}[h]
\centering
\includegraphics[scale=0.5]{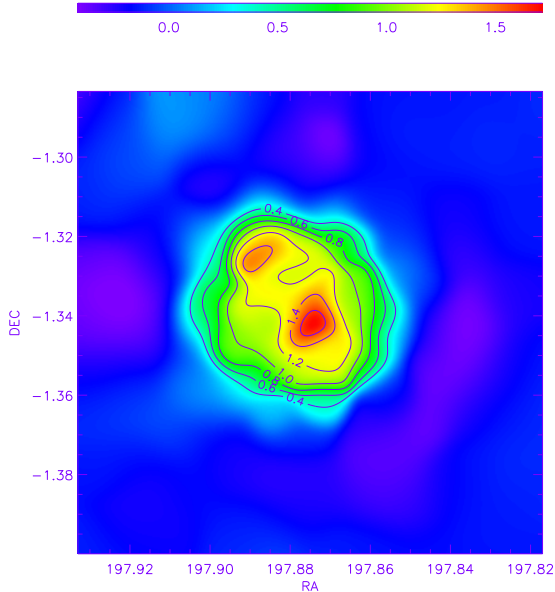}
\caption{ A Strong+Weak lensing mass reconstruction of A1689 using Particle Based Lensing. The contours of the field of view represent values of $\kappa$. The color-scale gives the value of $\kappa$. In this figure north is up and east is toward the left. Two clumps of dark matter are detected, the secondary clump is toward the North-East of the main clump.}
\label{fig:S+W}
\end{figure*}

\begin{figure}[h]
\centering
\includegraphics[scale=0.5]{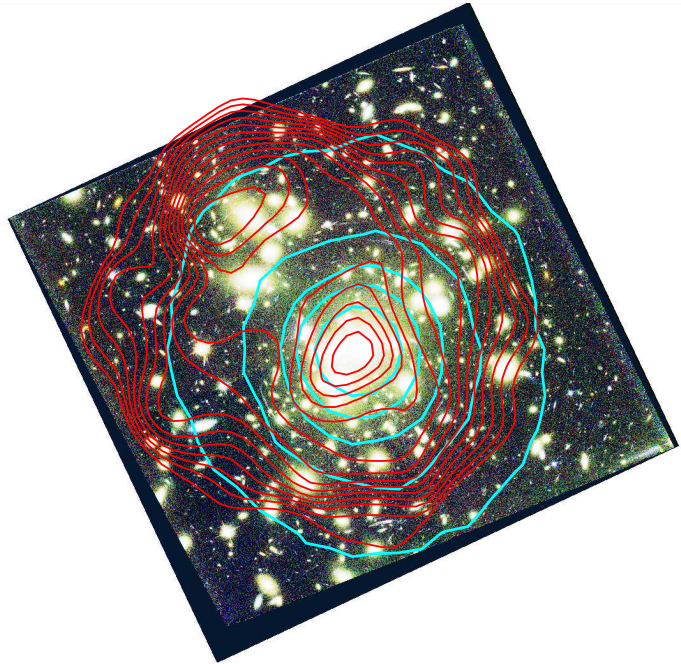}
\caption{Optical, X-ray and lensing composite image of A1689. The optical image is composed of ACS images from three wavelengths (F475W, F775W and F850LP). The red contours represent the dark matter distribution and the cyan contours represent the gas distribution. The optical image has a secondary clump of galaxies coincident with a sub-peak in the dark matter distribution in the North-East(NE) direction. The X-ray distribution is smoother, though there is an elongation in NE direction.}
\label{fig:comp}
\end{figure}

\begin{figure}[h]
\centering
\includegraphics[scale=0.4]{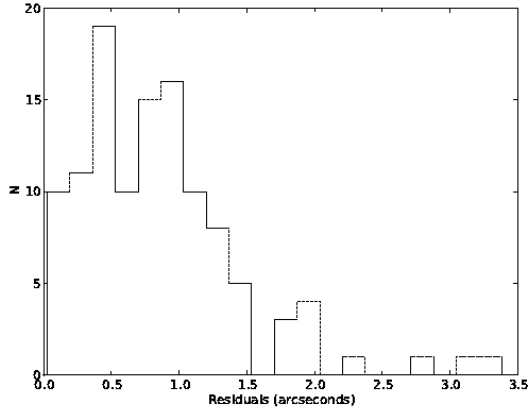}
\caption{Residuals from the fit to the strong lensing data. Most of the observables are fit within $1.5^{\prime\prime}$.  The mass is simultaneously fit to the strong and weak lensing data, thus not all strong lensing constraints are fit tightly. }
\label{fig:resi}
\end{figure}

We have applied PBL as described in \S~\ref{sec:ws} to the strong and weak lensing data for A1689. Figure~\ref{fig:S+W} shows the mass map for $7^\prime\times7^\prime=1570$ {kpc} $\times 1570$ {kpc} field of view centered on the cluster. The resolution of the mass map in the regions containing strong lensing data, i.e the central $4$ square arcminutes is $\sim17.5^{\prime\prime}$. The rest of the mass map has a resolution of $\sim50^{\prime\prime}$. This is necessary since the SUBARU data has $\sim12$ galaxies per square arcminutes. The higher resolution at the center of the map enables us to resolve more secondary structure.
 The advantage of computing the covariance matrix is that we are able to calculate physical parameters (with errors bars) related to the mass distribution.
The mass map in the left panel of Figure~\ref{fig:S+W} shows a secondary peak in the north-east direction to the cluster at a significance of $5\sigma$.
 Compared to previous strong lensing only mass reconstruction we are able to probe the mass distribution out to a larger radius. The presence of the secondary structure is revealed on addition of the strong lensing data. The optical image shows some second group of galaxies at that location. 
 In order to illustrate this we have provided a composite map of the lensing, X-ray and optical image (Figure~\ref{fig:comp}). The optical data is from HST/ACS composite of three wavelengths (F475W,  F775W and F850LP). We observe a secondary peak  at a distance of $180 \pm 7.5$ kpc North-East of the main dark matter peak. 

 In order to quantify these secondary structures, we use the centroid and the ellipticity and power ratio as a function of radial distance from the central peak. The centroid of the lensing mass distribution is $RA=197.873^o\pm0.008$, $DEC=-1.3437^o\pm0.0008$ and $30\pm7$ kpc away from the central BCG and does not change significantly with radial distance. The X-ray centroid is 25 kpc away from the BCG. The centroid for the gas at $RA=197.873^o$,  $DEC=-1.341^o$ and the dark matter are coincident within errors. The ellipticity of the dark matter distribution is consistently higher than the X-ray distribution.  It rises sharply within the first 200 kpc to asymptote to a value of $0.152\pm0.016$. In the inner 100 kpc ($28^{\prime\prime}$) the ellipticity is only sensitive to one peak and hence it is lower, as the secondary peak gets included in the calculation the ellipticity of the distribution rises. The X-ray distribution is smoother and rounder. This is consistent with the theoretical results  \cite[][see Figure 4 and Figure 5]{2003ApJ...585..151L} and simulations \citep{2001ApJ...558..520Y}. This suggests that the gas pressure is isotropic while the collisionless dark matter has anisotropic velocity ellipsoids.
  The position angle for both the X-ray and the lensing does not change significantly with radius. At 500 kpc the position angle is $11\pm5.7^{o}$ for the
S+W mass reconstruction and $13\pm0.67^o$ for the X-ray distribution. 
 
We also calculated the power ratios shown in  Figure~\ref{fig:pratio} for this cluster. The lens mass distribution has higher power than the  X-ray for the second (P2/P0) and third (P3/P0) powers. This implies that the dark matter distribution is clumpier than the gas distribution. The fourth power ratio (P4/P0) is dominated by noise and values represent an upper limit to the measurements. The cumulative power ratios decrease as we approach a radial distance of $1$ Mpc from the center since the substructure gets smeared out at large radius.  The zone of higher power represents the clumpy central core and the region of lower power the relaxed cluster at $1$ Mpc.  
 

  \begin{figure}[h]
\centering
\includegraphics[scale=0.4]{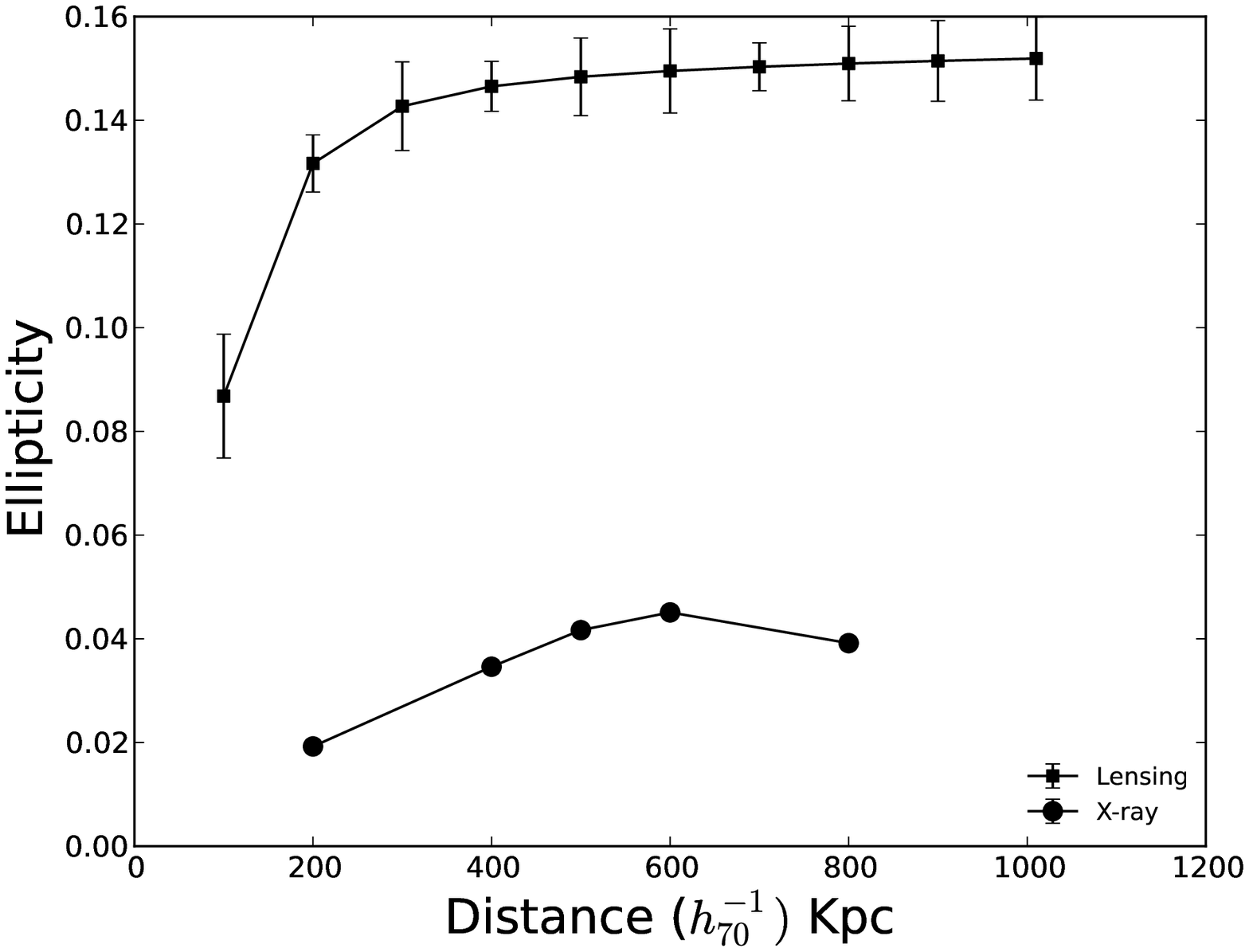}
\caption{Dependence of ellipticity of the dark matter (lensing) and gas (xray)  halo on radial distance. The ellipticity of the dark matter halo rises sharply in the inner $100-200$ kpc and asymptotes to a higher value. }
\label{fig:ellip}
\end{figure}

 \begin{figure}[h]
\centering
\includegraphics[scale=0.55]{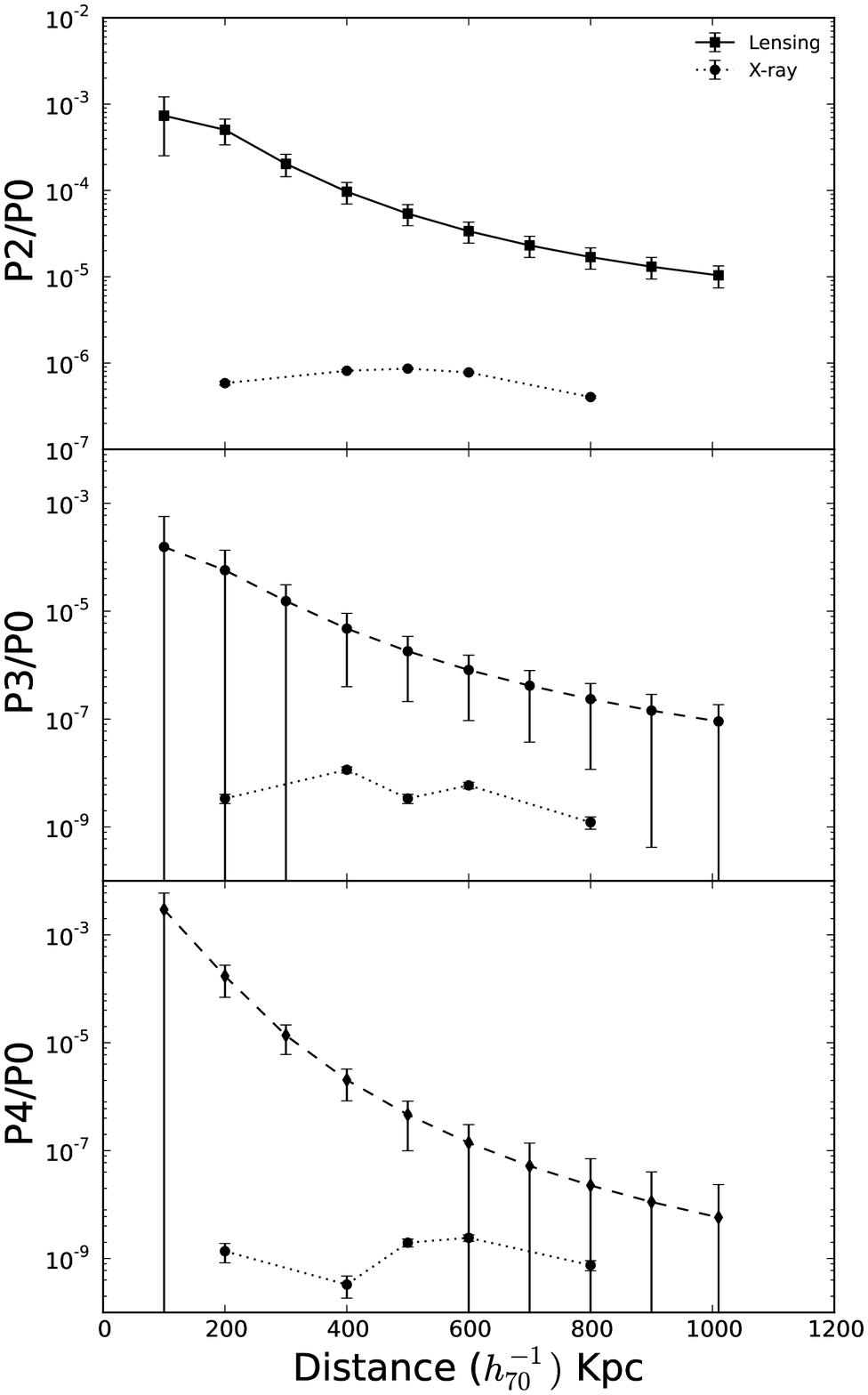}
\caption{Power ratio as a function of radial distance. The higher power in lensing mass distribution than in the X-ray distribution. The power also decreases with distance since the substructure dominates at the inner core.}
\label{fig:pratio}
\end{figure}

  
We have done all measurements from Strong+Weak lensing mass map. The dominating source of error in the mass map is the weak lensing uncertainties caused by the intrinsic shape noise in the background galaxies and their low number density. The error of the entire map is higher than a strong lensing only reconstruction of the central core.  The map also has variable resolution implying that structure is detected at a variable scale. This can be a possible source for systematic error.
If there are low mass structures present in the weak lensing regime the $50^{\prime\prime}$ resolution will smooth it away thus biasing the power ratio calculations. The effect of such systematics will be studied from simulations in future work. The effect of weak lensing dilution should also be low since we have removed object with very high ellipticity.
\subsection{NFW Fitting}
\begin{figure}[h]
\centering
\includegraphics[scale=0.5]{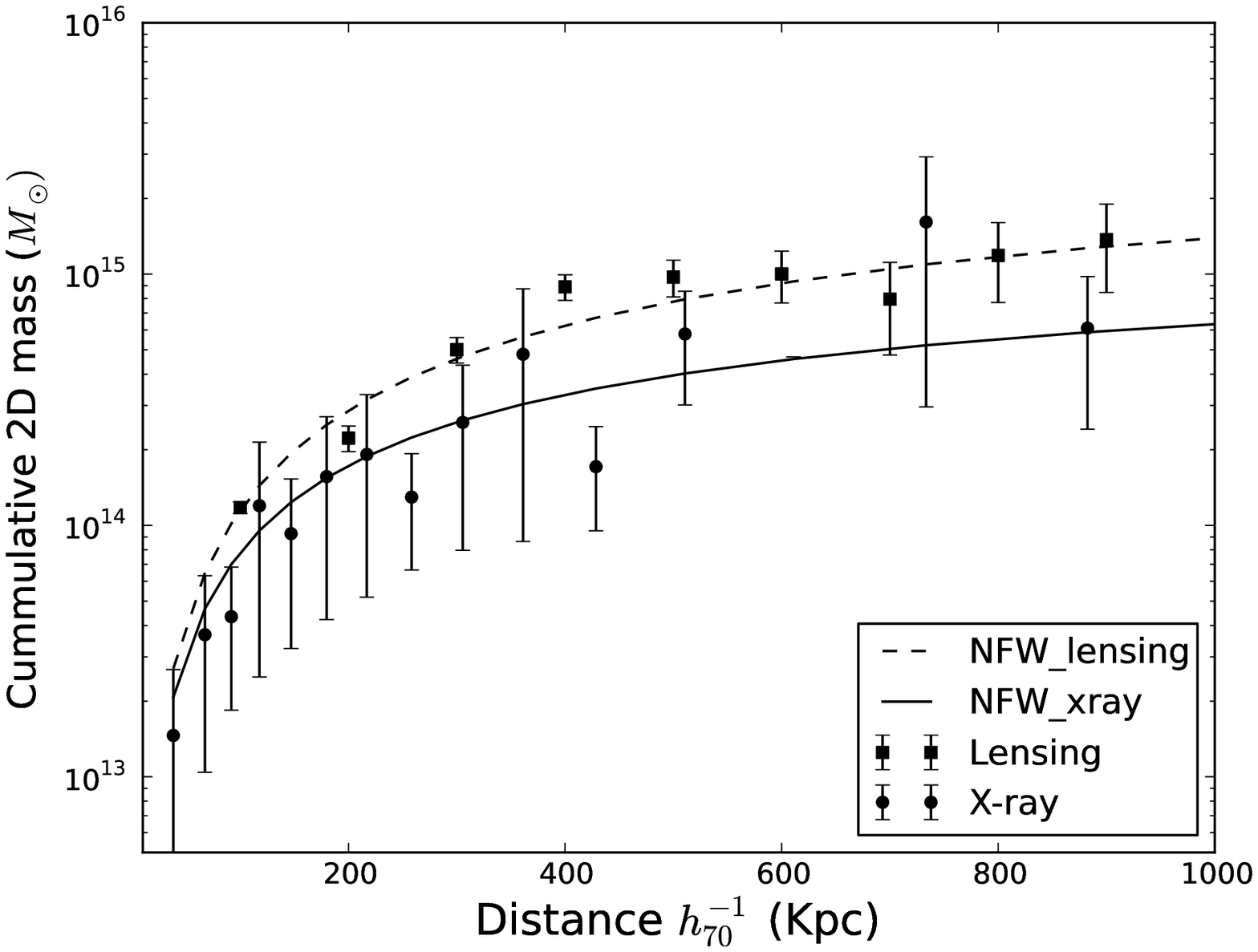}
\caption{ Comparison of the NFW fit between the Lensing and the X-ray data. The lensing mass marginally higher the X-ray mass.}
\label{fig:nfw}
\end{figure}

We calculate the cumulative mass derived from the Strong+Weak lensing mass map by integrating the mass within a given radial distance Figure~\ref{fig:nfw}. We fit an projected NFW \citep{2000ApJ...534...34W} profile to the projected mass and compare it with results from the X-ray measurements. The best fit values to the lensing results are shown by the dashed lines. We find the value of $r_{200}=1.773\pm 0.32$ and the concentration $c=5.56 \pm 1.26$. The mass enclosed within 1 Mpc is $(1.5\pm 0.33)\times10^{15}M_\odot$. The projected lensing is systematically higher than the X-ray mass calculated in \cite{2009ApJ...693.1570R}. Also the latest calibrations of Chandra data suggest a $10\%$ decrease in the measured X-ray mass. This is increase the discrepancy between X-ray and lensing masses.

\subsection{Comparison with previous results}
A1689 has been extensively studied multiple wavelengths. C10 summarize the most recent results on concentration in Table 2. They have made the highest resolution mass map of the inner $2^\prime \times2^\prime$ region using strong lensing only and obtained a clumpy distribution with zero residuals. Compared to the C10 mass map the PBL mass reconstruction has a lower resolution at the core  primarily to incorporate weak and strong lensing in the same mass map. The advantage of an S+W map is that we get a mass distribution out to a large radius. For A1689 we have made a twelve times twelve square arcminute field of view. Figure~\ref{fig:S+W} is a zoom into the central seven times seven square arcminute  region. The larger map has enabled us to calculate substructure parameters out to a large radius. The advantage of adding strong lensing to the weak lensing data is the detection of the secondary structure identified in the NE direction. This substructure is also detected by L07 and is coincident with secondary group of galaxies (Figure~\ref{fig:comp}) in the optical.  Presence of substructure in the NE direction has also been confirmed with 
velocity dispersion analysis of more than 500 cluster members suggests the existence of secondary structure that coincides with a group of galaxies with north-east to the central region \citep{2004ogci.conf..183C}. \cite{2009ApJ...693.1570R} have also detected substructure in the north-east direction. Recent SUZAKU observations in the X-rays \citep{2010ApJ...714..423K} suggests anisotropic gas and temperature distributions in the cluster outskirts correlated with the presence of the large scale structure galaxies in the photometric redshift slice centered around the cluster. Some studies have revealed a discrepancy between lensing and X-ray masses \citep{2009ApJ...701.1283P}, which have been addressed by triaxial modeling  of X-ray and strong lensing \citep{2011ApJ...729...37M} and by non-thermal gas pressure \citep{2010arXiv1002.4691M}. 
One of the peaks found by \cite{2010arXiv1009.1018L} is also coincident with the secondary structure in the NE direction.
\cite{2010A&A...518L..19H} have observed the A1689 field of view in the K-band and found that the light distribution is elongated in the $10-12^\prime$ scale. Valentina Vaccas et al. found that the substructure region has a lower radio emission than the overall cluster \citep{2011A&A...535A..82V}.

\section{Discussion and Future Work}
In this paper we have reconstructed the mass map of A1689 using Strong and Weak lensing data. This is the first non-parametric S+W mass map with a covariance matrix. The mass map of A1689 has a higher resolution at the cluster core where we have strong lensing data and it is smoother toward the periphery.  We have also measured the substructure parameters from lensing and X-ray distribution and found that the dark matter is clumpier than the gas distribution. The covariance calculation technique developed in this paper applies to grid based $\chi^2$ minimization methods for mass reconstruction as well. The substructure analysis can be applied to a sample of clusters at varying mass and redshift and test the predictions of $\Lambda$CDM cosmology. Additionally we can fit the dependence of scatter in the scaling relationships between mass and temperature on substructure from simulations \cite{2008ApJ...685..118V} and correct the observed scaling relations at different redshift.

The current scenario with observations is ideal for detailed study of cluster properties. There are several groups doing multiwavelength analysis of galaxy clusters at different redshift range.  The Cluster Lensing And Supernova survey with Hubble (CLASH,  $0.18<z<0.9 $, $ 5\times 10^{14} h^{-1} M_\odot<M<3\times 10^{15} h^{-1} M_\odot$.) is observing clusters to produce deep HST images similar to A1689. Future surveys like LSST and EUCLID will have lensing data for large parts of the sky spanning a wide redshift range will provide a cluster sample with well understood selection function. These future datasets will be optimal for the substructure analysis presented in this paper. 

\section{Acknowledgements}
SD acknowledges the support of LDRD program at Argonne National Laboratory. SD also acknowledges Marceau Limousin and Peter Schneider for useful discussions.
DMG  acknowledges support of NSF 0908307.
A.M. acknowledges support by Israel Science Foundation grant 823/09.
The Dark Cosmology Centre is funded by the Danish National Research Foundation.
KP acknowledges support from Instrument Center for Danish Astrophysics

\newcommand{\noopsort}[1]{}


\begin{thebibliography}{57}
\expandafter\ifx\csname natexlab\endcsname\relax\def\natexlab#1{#1}\fi

\bibitem[{{Allen}(1998)}]{1998MNRAS.296..392A}
{Allen}, S.~W. 1998, \mnras, 296, 392

\bibitem[{{Bah{\'e}} {et~al.}(2011){Bah{\'e}}, {McCarthy}, \&
  {King}}]{2011arXiv1106.2046B}
{Bah{\'e}}, Y.~M., {McCarthy}, I.~G., \& {King}, L.~J. 2011, ArXiv e-prints

\bibitem[{{Bardeau} {et~al.}(2005){Bardeau}, {Kneib}, {Czoske}, {Soucail},
  {Smail}, {Ebeling}, \& {Smith}}]{2005A&A...434..433B}
{Bardeau}, S., {Kneib}, J.-P., {Czoske}, O., {Soucail}, G., {Smail}, I.,
  {Ebeling}, H., \& {Smith}, G.~P. 2005, \aap, 434, 433

\bibitem[{{Bartelmann} {et~al.}(2006){Bartelmann}, {Doran}, \&
  {Wetterich}}]{2006A&A...454...27B}
{Bartelmann}, M., {Doran}, M., \& {Wetterich}, C. 2006, \aap, 454, 27

\bibitem[{{Broadhurst} {et~al.}(2005{\natexlab{a}}){Broadhurst}, {Takada},
  {Umetsu}, {Kong}, {Arimoto}, {Chiba}, \& {Futamase}}]{2005ApJ...619L.143B}
{Broadhurst}, T., {Takada}, M., {Umetsu}, K., {Kong}, X., {Arimoto}, N.,
  {Chiba}, M., \& {Futamase}, T. 2005{\natexlab{a}}, \apjl, 619, L143

\bibitem[{{Broadhurst} {et~al.}(2005{\natexlab{b}}){Broadhurst},
  {Ben{\'{\i}}tez}, {Coe}, {Sharon}, {Zekser}, {White}, {Ford}, {Bouwens},
  {Blakeslee}, {Clampin}, {Cross}, {Franx}, {Frye}, {Hartig}, {Illingworth},
  {Infante}, {Menanteau}, {Meurer}, {Postman}, {Ardila}, {Bartko}, {Brown},
  {Burrows}, {Cheng}, {Feldman}, {Golimowski}, {Goto}, {Gronwall}, {Herranz},
  {Holden}, {Homeier}, {Krist}, {Lesser}, {Martel}, {Miley}, {Rosati},
  {Sirianni}, {Sparks}, {Steindling}, {Tran}, {Tsvetanov}, \&
  {Zheng}}]{2005ApJ...621...53B}
{Broadhurst}, T., {et~al.} 2005{\natexlab{b}}, \apj, 621, 53

\bibitem[{{Buote} \& {Canizares}(1994)}]{buote1994}
{Buote}, D.~A., \& {Canizares}, C.~R. 1994, \apj, 427, 86

\bibitem[{{Buote} \& {Tsai}(1995)}]{1995ApJ...452..522B}
{Buote}, D.~A., \& {Tsai}, J.~C. 1995, \apj, 452, 522

\bibitem[{{Coe} {et~al.}(2010){Coe}, {Benitez}, {Broadhurst}, {Moustakas}, \&
  {Ford}}]{2010arXiv1005.0398C}
{Coe}, D., {Benitez}, N., {Broadhurst}, T., {Moustakas}, L., \& {Ford}, H.
  2010, ArXiv e-prints

\bibitem[{{Cypriano} {et~al.}(2001){Cypriano}, {Sodr{\'e}}, {Campusano},
  {Kneib}, {Giovanelli}, {Haynes}, {Dale}, \& {Hardy}}]{2001AJ....121...10C}
{Cypriano}, E.~S., {Sodr{\'e}}, L.~J., {Campusano}, L.~E., {Kneib}, J.-P.,
  {Giovanelli}, R., {Haynes}, M.~P., {Dale}, D.~A., \& {Hardy}, E. 2001, \aj,
  121, 10

\bibitem[{{Czoske}(2004)}]{2004ogci.conf..183C}
{Czoske}, O. 2004, in IAU Colloq. 195: Outskirts of Galaxy Clusters: Intense
  Life in the Suburbs, ed. {A.~Diaferio}, 183--187

\bibitem[{{Deb} {et~al.}(2010){Deb}, {Goldberg}, {Heymans}, \&
  {Morandi}}]{2010ApJ...721..124D}
{Deb}, S., {Goldberg}, D.~M., {Heymans}, C., \& {Morandi}, A. 2010, \apj, 721,
  124

\bibitem[{{Deb} {et~al.}(2008){Deb}, {Goldberg}, \&
  {Ramdass}}]{2008ApJ...687....1D}
{Deb}, S., {Goldberg}, D.~M., \& {Ramdass}, V.~J. 2008, \apj, 687, 39

\bibitem[{{Diego} {et~al.}(2005){Diego}, {Sandvik}, {Protopapas}, {Tegmark},
  {Ben{\'{\i}}tez}, \& {Broadhurst}}]{2005MNRAS.362.1247D}
{Diego}, J.~M., {Sandvik}, H.~B., {Protopapas}, P., {Tegmark}, M.,
  {Ben{\'{\i}}tez}, N., \& {Broadhurst}, T. 2005, \mnras, 362, 1247

\bibitem[{{Eke} {et~al.}(1996){Eke}, {Cole}, \& {Frenk}}]{1996MNRAS.282..263E}
{Eke}, V.~R., {Cole}, S., \& {Frenk}, C.~S. 1996, \mnras, 282, 263

\bibitem[{{Evrard} {et~al.}(2008){Evrard}, {Bialek}, {Busha}, {White}, {Habib},
  {Heitmann}, {Warren}, {Rasia}, {Tormen}, {Moscardini}, {Power}, {Jenkins},
  {Gao}, {Frenk}, {Springel}, {White}, \& {Diemand}}]{2008ApJ...672..122E}
{Evrard}, A.~E., {et~al.} 2008, \apj, 672, 122

\bibitem[{{Francis} {et~al.}(2009){Francis}, {Lewis}, \&
  {Linder}}]{2009MNRAS.393L..31F}
{Francis}, M.~J., {Lewis}, G.~F., \& {Linder}, E.~V. 2009, \mnras, 393, L31

\bibitem[{{Frenk} {et~al.}(1990){Frenk}, {White}, {Efstathiou}, \&
  {Davis}}]{1990ApJ...351...10F}
{Frenk}, C.~S., {White}, S.~D.~M., {Efstathiou}, G., \& {Davis}, M. 1990, \apj,
  351, 10

\bibitem[{{Gruen} {et~al.}(2011){Gruen}, {Bernstein}, {Lam}, \&
  {Seitz}}]{2011MNRAS.416.1392G}
{Gruen}, D., {Bernstein}, G.~M., {Lam}, T.~Y., \& {Seitz}, S. 2011, \mnras,
  416, 1392

\bibitem[{{Haines} {et~al.}(2010){Haines}, {Smith}, {Pereira}, {Egami},
  {Moran}, {Hardegree-Ullman}, {Rawle}, \& {Rex}}]{2010A&A...518L..19H}
{Haines}, C.~P., {Smith}, G.~P., {Pereira}, M.~J., {Egami}, E., {Moran}, S.~M.,
  {Hardegree-Ullman}, E., {Rawle}, T.~D., \& {Rex}, M. 2010, \aap, 518, L19+

\bibitem[{{Heymans} {et~al.}(2006){Heymans}, {Van Waerbeke}, {Bacon}, {Berge},
  {Bernstein}, {Bertin}, {Bridle}, {Brown}, {Clowe}, {Dahle}, {Erben}, {Gray},
  {Hetterscheidt}, {Hoekstra}, {Hudelot}, {Jarvis}, {Kuijken}, {Margoniner},
  {Massey}, {Mellier}, {Nakajima}, {Refregier}, {Rhodes}, {Schrabback}, \&
  {Wittman}}]{2006MNRAS.368.1323H}
{Heymans}, C., {et~al.} 2006, \mnras, 368, 1323

\bibitem[{{Heymans} {et~al.}(2008){Heymans}, {Gray}, {Peng}, {van Waerbeke},
  {Bell}, {Wolf}, {Bacon}, {Balogh}, {Barazza}, {Barden}, {B{\"o}hm},
  {Caldwell}, {H{\"a}u{\ss}ler}, {Jahnke}, {Jogee}, {van Kampen}, {Lane},
  {McIntosh}, {Meisenheimer}, {Mellier}, {S{\'a}nchez}, {Taylor}, {Wisotzki},
  \& {Zheng}}]{2008MNRAS.385.1431H}
---. 2008, \mnras, 385, 1431

\bibitem[{{Ho} {et~al.}(2006){Ho}, {Bahcall}, \& {Bode}}]{2006ApJ...647....8H}
{Ho}, S., {Bahcall}, N., \& {Bode}, P. 2006, \apj, 647, 8

\bibitem[{{Hoekstra} {et~al.}(1998){Hoekstra}, {Franx}, {Kuijken}, \&
  {Squires}}]{1998ApJ...504..636H}
{Hoekstra}, H., {Franx}, M., {Kuijken}, K., \& {Squires}, G. 1998, \apj, 504,
  636

\bibitem[{{Jeltema} {et~al.}(2005){Jeltema}, {Canizares}, {Bautz}, \&
  {Buote}}]{2005ApJ...624..606J}
{Jeltema}, T.~E., {Canizares}, C.~R., {Bautz}, M.~W., \& {Buote}, D.~A. 2005,
  \apj, 624, 606

\bibitem[{{Jeltema} {et~al.}(2008){Jeltema}, {Hallman}, {Burns}, \&
  {Motl}}]{2008ApJ...681..167J}
{Jeltema}, T.~E., {Hallman}, E.~J., {Burns}, J.~O., \& {Motl}, P.~M. 2008,
  \apj, 681, 167

\bibitem[{{Kaiser} {et~al.}(1995){Kaiser}, {Squires}, \&
  {Broadhurst}}]{1995ApJ...449..460K}
{Kaiser}, N., {Squires}, G., \& {Broadhurst}, T. 1995, \apj, 449, 460

\bibitem[{{Kawaharada} {et~al.}(2010){Kawaharada}, {Okabe}, {Umetsu},
  {Takizawa}, {Matsushita}, {Fukazawa}, {Hamana}, {Miyazaki}, {Nakazawa}, \&
  {Ohashi}}]{2010ApJ...714..423K}
{Kawaharada}, M., {et~al.} 2010, \apj, 714, 423

\bibitem[{{King} \& {Corless}(2007)}]{2007MNRAS.374L..37K}
{King}, L., \& {Corless}, V. 2007, \mnras, 374, L37

\bibitem[{{Kravtsov} {et~al.}(2006){Kravtsov}, {Vikhlinin}, \&
  {Nagai}}]{2006ApJ...650..128K}
{Kravtsov}, A.~V., {Vikhlinin}, A., \& {Nagai}, D. 2006, \apj, 650, 128

\bibitem[{{Lee} \& {Suto}(2003)}]{2003ApJ...585..151L}
{Lee}, J., \& {Suto}, Y. 2003, \apj, 585, 151

\bibitem[{{Lemze} {et~al.}(2009){Lemze}, {Broadhurst}, {Rephaeli}, {Barkana},
  \& {Umetsu}}]{2009ApJ...701.1336L}
{Lemze}, D., {Broadhurst}, T., {Rephaeli}, Y., {Barkana}, R., \& {Umetsu}, K.
  2009, \apj, 701, 1336

\bibitem[{{Leonard} {et~al.}(2010){Leonard}, {King}, \&
  {Goldberg}}]{2010arXiv1009.1018L}
{Leonard}, A., {King}, L.~J., \& {Goldberg}, D.~M. 2010, ArXiv e-prints

\bibitem[{{Limousin} {et~al.}(2007){Limousin}, {Richard}, {Jullo}, {Kneib},
  {Fort}, {Soucail}, {El{\'{\i}}asd{\'o}ttir}, {Natarajan}, {Ellis}, {Smail},
  {Czoske}, {Smith}, {Hudelot}, {Bardeau}, {Ebeling}, {Egami}, \&
  {Knudsen}}]{2007ApJ...668..643L}
{Limousin}, M., {et~al.} 2007, \apj, 668, 643

\bibitem[{{Luppino} \& {Kaiser}(1997)}]{1997ApJ...475...20L}
{Luppino}, G.~A., \& {Kaiser}, N. 1997, \apj, 475, 20

\bibitem[{{Mahdavi} {et~al.}(2007){Mahdavi}, {Hoekstra}, {Babul}, {Balam}, \&
  {Capak}}]{2007ApJ...668..806M}
{Mahdavi}, A., {Hoekstra}, H., {Babul}, A., {Balam}, D.~D., \& {Capak}, P.~L.
  2007, \apj, 668, 806

\bibitem[{{Miyazaki} {et~al.}(2002){Miyazaki}, {Komiyama}, {Sekiguchi},
  {Okamura}, {Doi}, {Furusawa}, {Hamabe}, {Imi}, {Kimura}, {Nakata}, {Okada},
  {Ouchi}, {Shimasaku}, {Yagi}, \& {Yasuda}}]{2002PASJ...54..833M}
{Miyazaki}, S., {et~al.} 2002, \pasj, 54, 833

\bibitem[{{Molnar} {et~al.}(2010){Molnar}, {Chiu}, {Umetsu}, {Chen}, {Hearn},
  {Broadhurst}, {Bryan}, \& {Shang}}]{2010arXiv1002.4691M}
{Molnar}, S.~M., {Chiu}, I., {Umetsu}, K., {Chen}, P., {Hearn}, N.,
  {Broadhurst}, T., {Bryan}, G., \& {Shang}, C. 2010, ArXiv e-prints

\bibitem[{{Morandi} {et~al.}(2011{\natexlab{a}}){Morandi}, {Limousin},
  {Rephaeli}, {Umetsu}, {Barkana}, {Broadhurst}, \&
  {Dahle}}]{2011MNRAS.416.2567M}
{Morandi}, A., {Limousin}, M., {Rephaeli}, Y., {Umetsu}, K., {Barkana}, R.,
  {Broadhurst}, T., \& {Dahle}, H. 2011{\natexlab{a}}, \mnras, 416, 2567

\bibitem[{{Morandi} {et~al.}(2011{\natexlab{b}}){Morandi}, {Pedersen}, \&
  {Limousin}}]{2011ApJ...729...37M}
{Morandi}, A., {Pedersen}, K., \& {Limousin}, M. 2011{\natexlab{b}}, \apj, 729,
  37

\bibitem[{{Navarro} {et~al.}(1997){Navarro}, {Frenk}, \&
  {White}}]{1997ApJ...490..493N}
{Navarro}, J.~F., {Frenk}, C.~S., \& {White}, S.~D.~M. 1997, \apj, 490, 493

\bibitem[{{Okabe} {et~al.}(2009){Okabe}, {Takada}, {Umetsu}, {Futamase}, \&
  {Smith}}]{2009arXiv0903.1103O}
{Okabe}, N., {Takada}, M., {Umetsu}, K., {Futamase}, T., \& {Smith}, G.~P.
  2009, ArXiv e-prints

\bibitem[{{Pedersen} \& {Dahle}(2007)}]{2007ApJ...667...26P}
{Pedersen}, K., \& {Dahle}, H. 2007, \apj, 667, 26

\bibitem[{{Peng} {et~al.}(2009){Peng}, {Andersson}, {Bautz}, \&
  {Garmire}}]{2009ApJ...701.1283P}
{Peng}, E., {Andersson}, K., {Bautz}, M.~W., \& {Garmire}, G.~P. 2009, \apj,
  701, 1283

\bibitem[{{Press} \& {Schechter}(1974)}]{1974ApJ...187..425P}
{Press}, W.~H., \& {Schechter}, P. 1974, \apj, 187, 425

\bibitem[{{Riemer-S{\o}rensen} {et~al.}(2009){Riemer-S{\o}rensen}, {Paraficz},
  {Ferreira}, {Pedersen}, {Limousin}, \& {Dahle}}]{2009ApJ...693.1570R}
{Riemer-S{\o}rensen}, S., {Paraficz}, D., {Ferreira}, D.~D.~M., {Pedersen}, K.,
  {Limousin}, M., \& {Dahle}, H. 2009, \apj, 693, 1570

\bibitem[{{Sarazin}(1988)}]{1988xrec.book.....S}
{Sarazin}, C.~L. 1988, {X-ray emission from clusters of galaxies}, ed.
  {Sarazin, C.~L.}

\bibitem[{{Sheth} \& {Tormen}(1999)}]{1999MNRAS.308..119S}
{Sheth}, R.~K., \& {Tormen}, G. 1999, \mnras, 308, 119

\bibitem[{{Spinelli} {et~al.}(2011){Spinelli}, {Seitz}, {Lerchster},
  {Brimioulle}, \& {Finoguenov}}]{2011arXiv1111.0989S}
{Spinelli}, P.~F., {Seitz}, S., {Lerchster}, M., {Brimioulle}, F., \&
  {Finoguenov}, A. 2011, ArXiv e-prints

\bibitem[{{Vacca} {et~al.}(2011){Vacca}, {Govoni}, {Murgia}, {Giovannini},
  {Feretti}, {Tugnoli}, {Verheijen}, \& {Taylor}}]{2011A&A...535A..82V}
{Vacca}, V., {Govoni}, F., {Murgia}, M., {Giovannini}, G., {Feretti}, L.,
  {Tugnoli}, M., {Verheijen}, M.~A., \& {Taylor}, G.~B. 2011, \aap, 535, A82

\bibitem[{{Ventimiglia} {et~al.}(2008){Ventimiglia}, {Voit}, {Donahue}, \&
  {Ameglio}}]{2008ApJ...685..118V}
{Ventimiglia}, D.~A., {Voit}, G.~M., {Donahue}, M., \& {Ameglio}, S. 2008,
  \apj, 685, 118

\bibitem[{{Williams} {et~al.}(1999){Williams}, {Navarro}, \&
  {Bartelmann}}]{1999ApJ...527..535W}
{Williams}, L.~L.~R., {Navarro}, J.~F., \& {Bartelmann}, M. 1999, \apj, 527,
  535

\bibitem[{{Wright} \& {Brainerd}(2000)}]{2000ApJ...534...34W}
{Wright}, C.~O., \& {Brainerd}, T.~G. 2000, \apj, 534, 34

\bibitem[{{Yagi} {et~al.}(2002){Yagi}, {Kashikawa}, {Sekiguchi}, {Doi},
  {Yasuda}, {Shimasaku}, \& {Okamura}}]{2002AJ....123...66Y}
{Yagi}, M., {Kashikawa}, N., {Sekiguchi}, M., {Doi}, M., {Yasuda}, N.,
  {Shimasaku}, K., \& {Okamura}, S. 2002, \aj, 123, 66

\bibitem[{{Yoshikawa} {et~al.}(2001){Yoshikawa}, {Taruya}, {Jing}, \&
  {Suto}}]{2001ApJ...558..520Y}
{Yoshikawa}, K., {Taruya}, A., {Jing}, Y.~P., \& {Suto}, Y. 2001, \apj, 558,
  520

\bibitem[{{Zitrin} {et~al.}(2010{\natexlab{a}}){Zitrin}, {Broadhurst}, {Coe},
  {Liesenborgs}, {Benitez}, {Rephaeli}, {Ford}, \&
  {Umetsu}}]{2010arXiv1009.3936Z}
{Zitrin}, A., {Broadhurst}, T., {Coe}, D., {Liesenborgs}, J., {Benitez}, N.,
  {Rephaeli}, Y., {Ford}, H., \& {Umetsu}, K. 2010{\natexlab{a}}, ArXiv
  e-prints

\bibitem[{{Zitrin} {et~al.}(2010{\natexlab{b}}){Zitrin}, {Broadhurst},
  {Umetsu}, {Rephaeli}, {Medezinski}, {Bradley}, {Jim{\'e}nez-Teja},
  {Ben{\'{\i}}tez}, {Ford}, {Liesenborgs}, {de Rijcke}, {Dejonghe}, \&
  {Bekaert}}]{2010MNRAS.408.1916Z}
{Zitrin}, A., {et~al.} 2010{\natexlab{b}}, \mnras, 408, 1916

\end{thebibliography}
\end{document}